\documentclass[5p,sort&compress]{elsarticle}
\makeatletter
\def\ps@pprintTitle{%
 \let\@oddhead\@empty
 \let\@evenhead\@empty
 \def\@oddfoot{}%
 \let\@evenfoot\@oddfoot}
\makeatother
\usepackage{graphicx}
\usepackage{tabularx}
\usepackage{makeidx}
\usepackage{epstopdf}
\usepackage{epsfig}
\usepackage{subfig}
\usepackage{colortbl}
\usepackage{colordvi}
\usepackage{verbatim}
\usepackage{amsmath, amsthm}
\usepackage{enumerate}
\usepackage{wrapfig}
\usepackage[colorlinks=true,citecolor=blue,urlcolor=blue,linkcolor=blue]{hyperref}
\usepackage{dcolumn}
\usepackage{amssymb,mathrsfs}
\usepackage{widetext}
\usepackage{bm}

\journal{Physics Letters B}
\DeclareMathOperator{\sech}{sech}
\def\beq{\begin{equation}}
\def\eeq{\end{equation}}
\date{}

\begin{document}
\title{Greybody factor and sparsity of Hawking radiation from a charged spherical black hole with scalar hair}%

\author{Avijit Chowdhury}
\ead{ac13ip001@iiserkol.ac.in}

\author{Narayan Banerjee}
\ead{narayan@iiserkol.ac.in}

\address{IISER Kolkata, Mohanpur Campus, Mohanpur, Nadia, 741246, India}
\bibliographystyle{apsrev4-1}

\begin{abstract}
The ‘no-hair’ conjecture claims that for a spherically symmetric black hole, only the information regarding the mass and charge of the black hole is available to an external observer. However, there are numerous counterexamples to the ’no-hair’ conjecture. In this work, we consider a particular counter-example to the ’no-hair’ conjecture in (3+1) dimensions, namely, a static spherically symmetric charged black hole with a scalar hair. We provide semi-analytic bounds on the greybody factors and study the sparsity of Hawking radiation of mass-less uncharged scalar fields. Our results show that the scalar and electric charges contribute oppositely to the greybody factor and the sparsity of the Hawking radiation cascade. Also, the greybody factor decreases and the Hawking emission spectra become more sparse with the reduction in the black hole (ADM) mass.
\end{abstract}

\maketitle

\section{Introduction}

In 1974, Hawking\cite{hawking_CMP_1975} showed that a black hole behaves `almost' like a black body, spontaneously emitting particles at a temperature proportional to its surface gravity. However, to an asymptotic observer the Hawking emission spectrum is not an exact black-body-like Planckian distribution. The geometry interpolating between the black hole horizon and the asymptotic observer allows only a fraction of the emitted radiation to reach the asymptotic observer. This deviation of the Hawking emission spectrum from the perfect black body spectrum is described in terms of the greybody factor~\cite{page_PRD_1976,page_PRD2_1976}. Another important aspect of the Hawking radiation flow is its sparsity \cite{gray_CQG_2016, hod_PLB_2016, hod_EPJC_2015, miao_PLB_2017,schuster_thesis, bibhas_IJMPA_2017}. The Hawking emission is known to be extremely sparse, that is the average time gap between emission of successive Hawking quanta is large compared to the characteristic time-scale of individual Hawking emission.

In this work, we consider an electrically charged black hole  with an associated scalar hair, known in literature as the scalar-hairy-Reissner-Nordstr\"{o}m (scalar-hairy RN) black hole~\cite{astorino_PRD_2013}. The additional scalar field is conformally coupled to the Einstein-Maxwell gravity. The scalar hair manifests itself as an additive correction parameter (termed as the `scalar charge') to the square of  the electric charge in the standard Reissner-Nordstr\"{o}m (RN) metric and modifies the physics of the same. The quasinormal modes of the scalar-hairy RN black hole studied in Ref.\cite{chowdhury_EPJC_2018} shows nontrivial signature of the scalar hair in the quasinormal spectrum. The scalar-hairy RN spacetime is also found to be superradiantly stable\cite{chowdhury_GERG_2019} against perturbation by massive charged scalar fields. The Hawking emission of charged particles from the scalar-hairy RN black hole using the tunneling formalism was studied in Ref.\cite{chowdhury_EPJC_2019}. The total change in entropy of the black hole due to the emission of massive charged particles was found to  contain an additional frequency-dependent contribution due to the black hole scalar charge. The study also highlighted the nontrivial dependence of the black hole temperature on the scalar charge and the lowering of the maximum allowed charge-mass ratio of the emitted particles with the black hole scalar charge. So, one naturally asks, how the scalar hair affects the black hole greybody factor and the sparsity of the Hawking radiation? 

Earlier, there have been attempts to study the sparsity and greybody factor of Hawking radiation from higher dimensional ($D>4$) black holes with scalar hair (of a different nature than that considered in this paper, see Ref.\cite{miao_PLB_2017} for example), however, the (3+1) dimensional scalar-hairy RN black hole is still unexplored. This paper attempts to fill this gap and provide a cohesive understanding of the behaviour of the black hole greybody factor and the sparsity of Hawking radiation flow from the scalar-hairy RN black hole.

In the present work, we specifically consider the emission of massless uncharged scalar particles from the scalar-hairy RN black hole. We observe that unlike the electric charge which reduces the greybody factor, the scalar charge enhances the same. The scalar charge also reduces the sparsity of the Hawking cascade whereas the electric charge enhances it. With the (ADM) mass of the black hole the greybody factor increases and the sparsity of the Hawking emission cascade decreases.
 
The paper is organized as follows. We start with a brief review of the scalar-hairy RN black hole in section~\ref{sec:shRN}. In section~\ref{sec:hawking_rad}, we calculate the greybody factor and the sparsity of Hawking radiation of massless uncharged scalar particles from the scalar-hairy RN black hole. Finally, in section~\ref{sec:summary}, we provide a brief summary and discussion.  

\section{ Review of the scalar-hairy RN black hole} \label{sec:shRN}

We start with a very simple model available in literature \cite{astorino_PRD_2013,chowdhury_EPJC_2018, chowdhury_GERG_2019, chowdhury_EPJC_2019} in which general relativity is coupled to electromagnetic field $F^{\mu \nu}$ and conformally coupled to a scalar field $\psi$. The corresponding action reads,
\begin{equation}\label{eq_action}
\begin{split}
I=\frac{1}{16\pi }\int d^4 x\sqrt{-g}\left[ R-F_{\mu \nu}F^{\mu \nu}\right.\\ \left.-8\pi \left(\bigtriangledown_\mu \psi \bigtriangledown^\mu \psi+\frac{R}{6}\psi^2\right)\right].
\end{split}
\end{equation}
In this action and the following analysis we use the units in which $G=c=k_B=\hbar=1$.

The above model admits an electrically charged, static, spherically symmetric black hole endowed with a scalar hair~\cite{astorino_PRD_2013},
\begin{equation}\label{eq_metric}
ds^2=-f\left(r \right)dt^2+{f\left(r \right)}^{-1}dr^2+r^2 \left( d\theta^2+\sin^2{\theta} d\phi^2 \right),
\end{equation}
with 
\begin{equation}
\label{eq_f(r)}
f\left( r \right)=\left(1-\frac{2M}{r}+\frac{e^2+s}{r^2}\right)\quad \mbox{and} \quad \psi=\pm \sqrt{\frac{6}{8 \pi }}\sqrt{\frac{s}{s+e^2}}~,
\end{equation}
where $M$ and $e$ are respectively the mass parameter and electric charge of the black hole and $s$ is the scalar charge. The scalar field $\psi$ can survive even in the absence of the electromagnetic field and contributes non-trivially to the geometry. The scalar hair is thus a `primary hair'. The inner and outer horizons of the scalar-hairy RN black hole are located at
\begin{equation}\label{eq_r+-}
r_\pm=M \pm \sqrt{M^2-e^2-s}~.
\end{equation}
The total energy-momentum tensor due to the scalar and electromagnetic fields,
\begin{equation}\label{eq_stressenergy}
T^\mu_\nu=\frac{e^2+s}{r^4} diag\left(-1,-1,1,1\right),
\end{equation}
satisfies both the dominant and strong energy conditions for $s>-e^2$.  Throughout this work we will consider $s>0$.

The presence of the scalar hair also modifies the ADM mass of the black hole,
\begin{equation}\label{eq_MADM}
M_{ADM}=\frac{M}{1+s/e^2},
\end{equation}
from the standard Reissner-Nordstr\"{o}m case $\left(M_{ADM}=M\right)$ and gives rise to a Hawking temperature of 
\begin{equation}\label{eq_TBH}
\begin{split}
&T_{BH}=\frac{r_+ - r_-}{4 \pi r_+^2}\\
&=\frac{\sqrt{M_{ADM}^2 \left(\frac{s}{e^2}+1\right)^2-e^2-s}}{2 \pi  \left(\sqrt{M_{ADM}^2 \left(\frac{s}{e^2}+1\right)^2-e^2-s}+M_{ADM} \left(\frac{s}{e^2}+1\right)\right)^2}~.
\end{split}
\end{equation}


\section{Hawking emission of uncharged particles}\label{sec:hawking_rad}
The energy emitted per unit time by a black hole at temperature $T_{BH}$ with frequency $\omega$ in the momentum interval $d^3\overrightarrow{k}$
is given by\cite{miao_PLB_2017,gray_CQG_2016}
\begin{equation}
\frac{dE(\omega)}{dt}= \sum_l T_l (\omega)\frac{\omega}{e^{\omega/T_{BH}}-1} \hat{k} \cdot  \hat{n}~  \frac{d^3 k ~dA}{(2\pi)^3},
\end{equation}
where $\hat{n}$ is the unit normal to the surface element $dA$, $l$ is the angular momentum quantum number and $T_l(\omega)$ is the frequency dependent greybody factor. For massless particles $\big|\overrightarrow{k}\big|=\omega$. Integrating over the finite surface area $A$, we get the total power of Hawking radiation  as,
\begin{equation}\label{eq_P}
P=\sum_l \int_{0}^{\infty} P_l\left(\omega\right) d\omega,
\end{equation}
where
\begin{equation}\label{eq_Pl}
P_l\left(\omega\right)=\frac{A}{8\pi^2} T_l(\omega)\frac{\omega^3}{e^{\omega/T_{BH}}-1} 
\end{equation}
is the power emitted per unit frequency in the $l^{th}$ mode. The area $A$ is usually a multiple of the horizon area. For Schwarzschild black hole, $A$ is taken to be  $(27/4 )$ times the horizon area. This value of the effective surface area ensures that in the limit of vanishing electric charge of the black hole, the low-frequency results smoothly match with the high-frequency results. Here, we will consider the area $A$ to be the horizon area $A_H$, where $A_H= 4 \pi r_+^2$, as this does not affect the qualitative behaviour of the results.

The excitation of massless uncharged scalar fields around the scalar-hairy RN black hole is governed by the Klein-Gordon equation,
\begin{equation}\label{eq_KG}
\square \Phi _{l m}\left(t,r,\theta,\phi\right)=0
\end{equation} 
Decomposing the scalar field as
\begin{equation}
\Phi_{lm}\left(t,r,\theta,\phi\right)=e^{-i\omega t}S_{lm}\left(\theta\right)R_{lm}\left(r\right)e^{i m \phi},
\end{equation}
and using separation of variables, we get the radial Klein-Gordon equation in the tortoise coordinate $r_*$ \cite{chowdhury_EPJC_2018, chowdhury_GERG_2019}, as,
\begin{equation}\label{eq_KG}
\frac{d^2 R_{lm}(r)}{d r_*^2}+\left( \omega^2-V_{eff}(r) \right)R_{lm}(r)=0,
\end{equation}
where
\begin{equation}\label{eq_Veff_unchrgd}
V_{eff}=f(r)\left(-\frac{2\left( e^2+s \right)}{r^4} +\frac{2M}{r^3}+\frac{l(l+1)}{r^2}\right)
\end{equation}
is the effective potential, $\omega$ is the conserved frequency, $l$ is the spherical harmonic index and $m$ ($-l\leq m\leq l$) is the azimuthal harmonic index.
The tortoise coordinate, $r_*$,  defined by  $dr_*=dr/f(r)$, maps the semi-infinite region $\left[ \left.r_+,\infty \right)\right.$ to $(-\infty, \infty)$.
The effective potential $V_{eff}$, vanishes both at the horizon and near spatial infinity. Thus, the solution at the horizon consists only of ingoing modes whereas near infinity, it consists of both ingoing and outgoing modes. 

A fraction of the radiation emitted by the black hole is reflected back by the effective potential while the remaining is transmitted out. The greybody factor measures the transmission probability of the outgoing Hawking quanta to reach future infinity without being back-scattered by this effective potential.


\subsection{Bounds on the greybody factor} \label{subsec:bound}
There are various methods in literature \cite{cardoso_PRL_2006,cardoso_JHEP_2006,harmark_ATMP_2010,grain_PRD_2005,fernando_GERG_2005,ida_PRD_2003,cvetic_FDP_2000,cvetic_PRD_1998,klebanov_NPB_1997,neitzke_ARXIV_2003,motl_ATMP_2003} to estimate the greybody factors, however,  derivation of exact analytical expression of greybody factor is limited only to very few cases.  Following Refs.\cite{visser_PRA_1999, boonserm_AP_2008, boonserm_thesis,boonserm_PRD_2008, ngampitipan_JPCS_2013,boonserm_JHEP_2014, miao_PLB_2017}, in this work, we will provide rigorous  bounds on the greybody factors of the scalar-hairy RN black hole. 
The general bounds on the greybody factor, as proposed by Visser\cite{visser_PRA_1999} is given by,
\begin{equation}\label{eq_bound}
T_l(\omega)\geq \sech^2\left\{ \int_{-\infty}^{\infty} \vartheta dr_* \right \}
\end{equation}
where \begin{equation}\label{eq_bound1}
\vartheta=\frac{\sqrt{\left[h'(r)\right]^2+\left[ \omega^2-V_{eff}-h(r)^2 \right]^2}}{2h(r)}.
\end{equation}
The arbitrary function $h(r)$ has to be positive definite everywhere and satisfy the boundary condition, $h(\infty)=h(r_+)=\omega$ for the bound~\eqref{eq_bound} to hold. A particularly simple choice of $h(r)$ for the present case is 
\begin{equation}\label{eq_h_unchrgd}
h(r)=\omega.
\end{equation}
Substituting Eq.\eqref{eq_h_unchrgd} in Eq.\eqref{eq_bound1} and using the definition of $r_*$, we get
\begin{equation}\label{eq_bound2}
\int_{-\infty}^{\infty} \vartheta dr_*=\int_{r_+}^{\infty} \frac{V_{eff}}{2 \omega f(r)}  dr.
\end{equation}
Eq.\eqref{eq_bound} in conjunction with  Eqs.\eqref{eq_f(r)},\eqref{eq_r+-} \eqref{eq_Veff_unchrgd} and \eqref{eq_bound2} yields a relatively simple expression for the lower bound of the greybody factor,
\begin{equation}\label{eq_tl_unchrgd_rh}
T_l(\omega)\geq \sech^2\left\{\frac{1+2l(l+1)}{4r_+ \omega}-\frac{e^2+s}{12r_+^3 \omega} \right\}.
\end{equation}
Using the expression of the horizon radius, $r_+$ (Eq\eqref{eq_r+-}), in terms of the ADM mass  (Eq.\eqref{eq_MADM}), we finally obtain, 
\begin{widetext}
	\begin{equation}\label{eq_tl_unchrgd_Madm}
	\begin{split}	
	&T_l(\omega)\geq \sech^2\left\lbrace\frac{3  s M_{ADM}  \left(\sqrt{ M_{ADM} ^2 \left(\frac{s}{e^2}+1\right)^2-e^2-s}+ M_{ADM}  \left(\frac{s}{e^2}+1\right)\right)-2 e^2\left(e^2 + s\right)}{6 e^2 \omega  \left(\sqrt{ M_{ADM} ^2 \left(\frac{s}{e^2}+1\right)^2-e^2-s}+ M_{ADM}  \left(\frac{s}{e^2}+1\right)\right)^3}\right.+\\
	&\left.\frac{3 e^2 \left(\sqrt{ M_{ADM} ^2 \left(\frac{s}{e^2}+1\right)^2-e^2-s}+ M_{ADM}  \left(\frac{s}{e^2}+1\right)\right) \left(l (l+1) \left(\sqrt{ M_{ADM} ^2 \left(\frac{s}{e^2}+1\right)^2-e^2-s}+ M_{ADM}  \left(\frac{s}{e^2}+1\right)\right) M_{ADM} \right)}{6 e^2 \omega  \left(\sqrt{ M_{ADM} ^2 \left(\frac{s}{e^2}+1\right)^2-e^2-s}+ M_{ADM}  \left(\frac{s}{e^2}+1\right)\right)^3}\right\rbrace .
	\end{split}
	\end{equation} 
\end{widetext}

To better understand the dependence of the greybody factor and the power spectrum on the black hole scalar and electric charges and the black hole mass, we plot the lower bound of the greybody factor~\eqref{eq_tl_unchrgd_Madm} and the power spectrum\eqref{eq_Pl}  for different values of $s$, $e$ and $M_{ADM}$.
\begin{figure}[!t]
\centering
\includegraphics[width=0.9\linewidth]{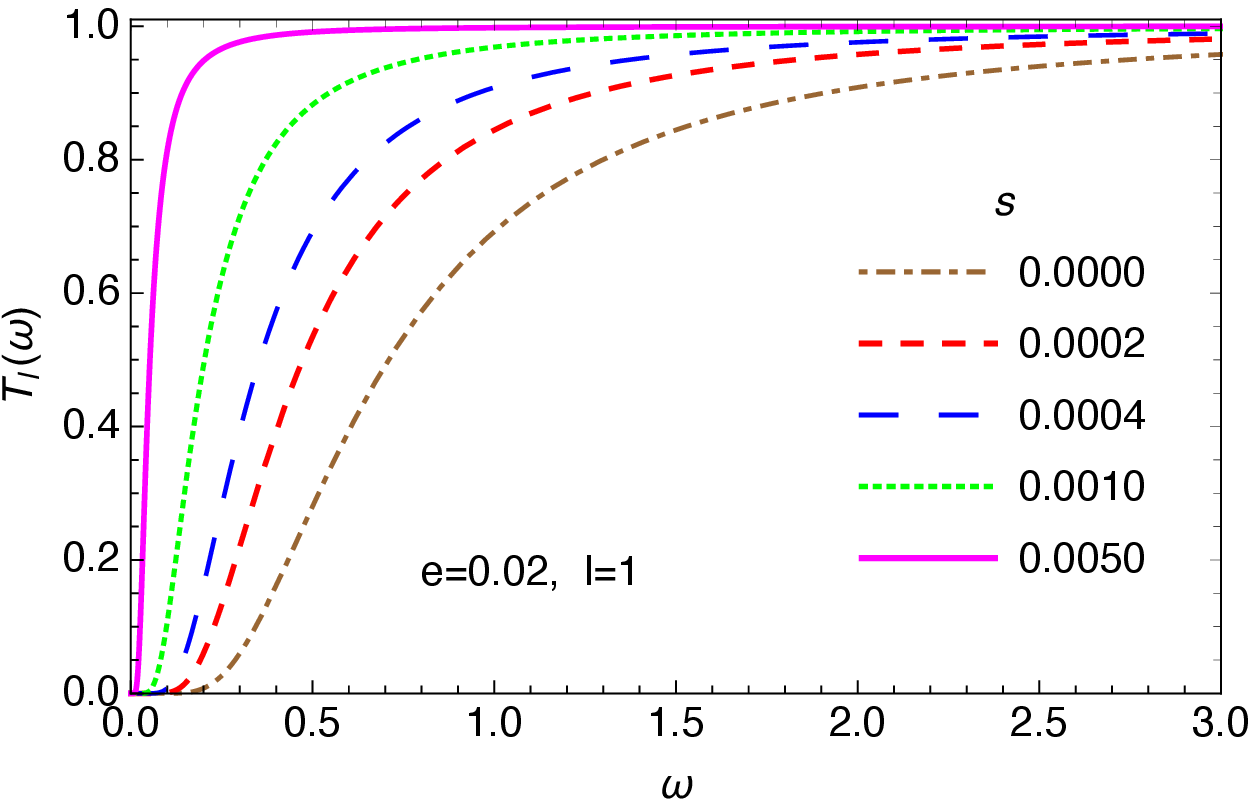}\\
\includegraphics[width=0.9\linewidth]{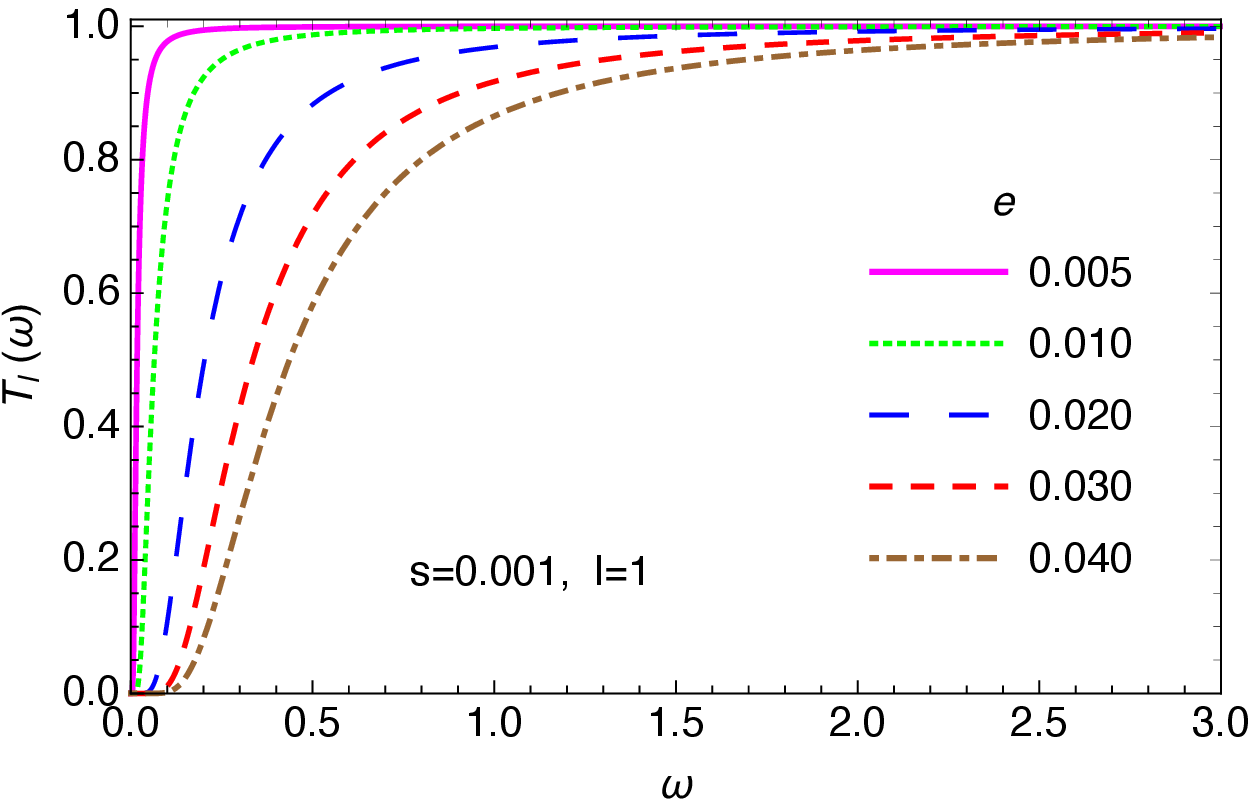}\\
\includegraphics[width=0.9\linewidth]{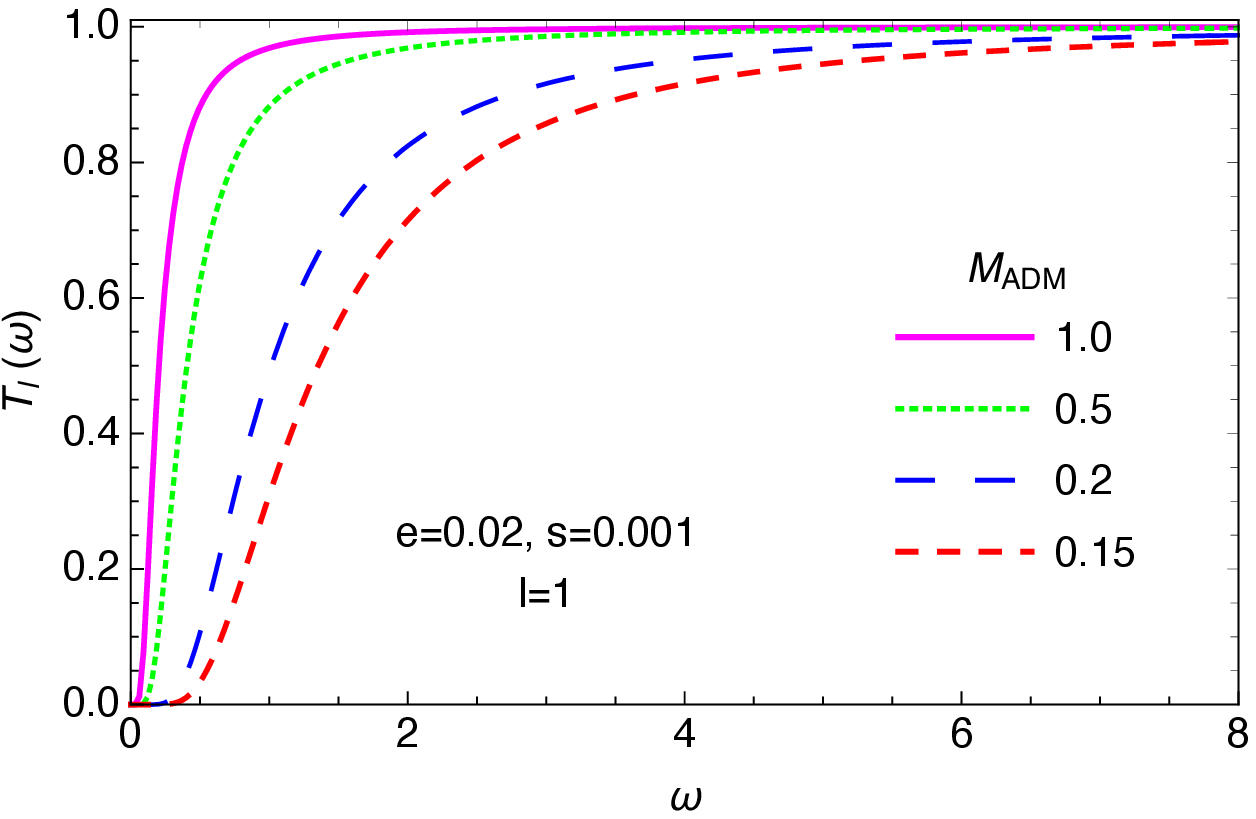}
\caption{{\bf (Top)} Plot of lower bound of the greybody factor $T_l(\omega)$ with $\omega$ for different values of the scalar charge $s$ with $e=0.02$, $l=1$ and $M_{ADM}=1$. {\bf (Middle)} Plot of lower bound of the greybody factor $T_l(\omega)$ with $\omega$ for different values of the electric charge $e$ with $s=0.001$, $l=1$ and $M_{ADM}=1$. {\bf (Bottom)} Plot of lower bound of the greybody factor $T_l(\omega)$ with $\omega$ for different values of  $M_{ADM}$ with $e=0.02$,  $s=0.001$ and $l=1$.}\label{fig:1}
\end{figure}
\begin{figure}[t!]
\includegraphics[width=0.95\linewidth]{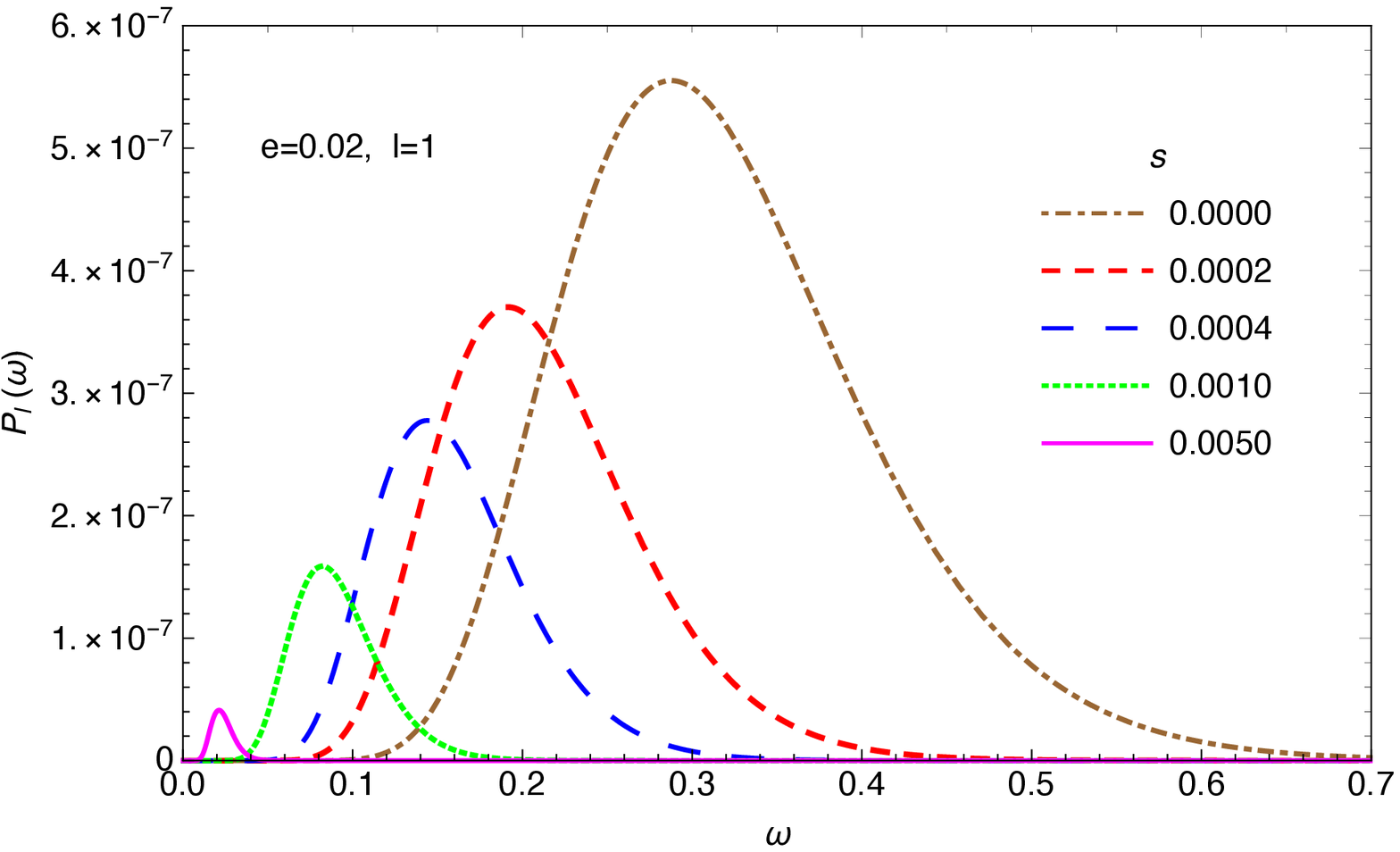}\\
\includegraphics[width=0.95\linewidth]{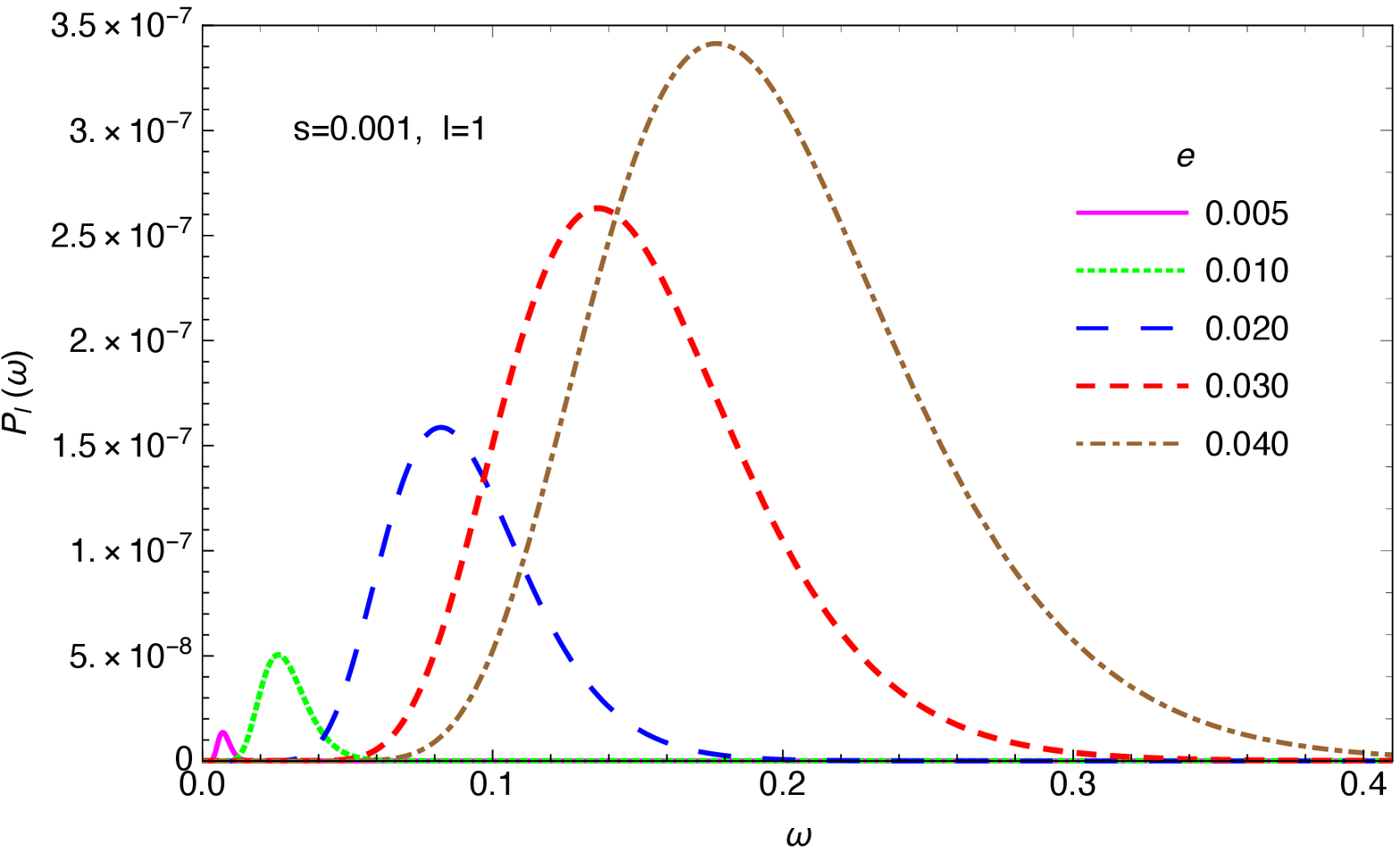}\\
\includegraphics[width=0.95\linewidth]{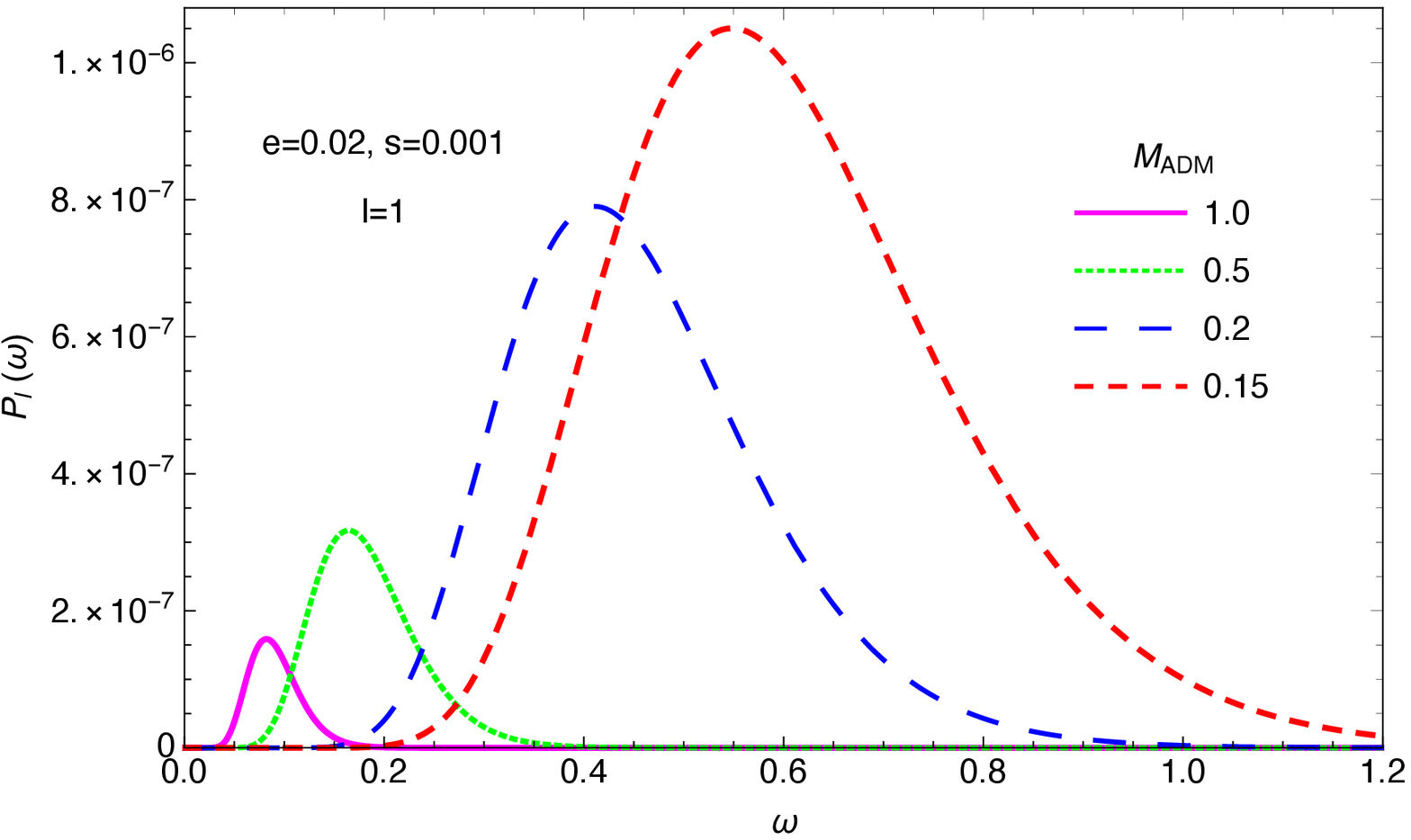}
\caption{{\bf (Top)} Plot of the power spectrum $P_l(\omega)$ with $\omega$ for different values of the scalar charge $s$ with $e=0.02$, $l=1$ and $M_{ADM}=1$. {\bf (Middle)} Plot of the power spectrum $P_l(\omega)$ with $\omega$ for different values of the electric charge $e$ with $s=0.001$, $l=1$ and $M_{ADM}=1$. {\bf (Bottom)} Plot of the power spectrum $P_l(\omega)$ with $\omega$ for different values of  $M_{ADM}$ with $e=0.02$,  $s=0.001$ and $l=1$.}\label{fig:2}
\end{figure}
We observe from the top and middle panels of Fig.\ref{fig:1}, that for a fixed ADM mass and spheroidal harmonic index, the lower bound of the greybody factor increases with the increase in the black hole scalar charge for a given value of the black hole electric charge whereas,  for fixed non zero values of the scalar charge, it decreases with increasing values of the electric charge.  The bottom panel of Fig.\ref{fig:1} shows that for fixed values of the black hole electric and scalar charges, the lower bound of the greybody factor decreases with the decrease in the black hole ADM mass.  In all cases, however, the far away observer misses more of the lower frequency contribution than the higher frequencies.  From the top and middle panels of Fig.\ref{fig:2}, we see that for a fixed ADM mass and spheroidal harmonic index, the peak of the power spectrum decreases and shifts towards lower frequencies as the scalar charge increases for a given value of the electric charge of the black hole, whereas, for a given non-zero scalar charge, it increases and shifts towards higher frequencies with the increase in  the electric charge. The bottom panel of Fig.\ref{fig:1} shows that for fixed values of the black hole electric and scalar charges, the peak of the power spectrum increases and shifts towards higher frequencies with the decrease in the black hole ADM mass.

\subsection{Sparsity of Hawking radiation}\label{subsec:sparsity}

To discuss the sparsity of Hawking radiation quantitatively, we use the dimensionless parameter $\eta$, defined in literature \cite{gray_CQG_2016, hod_PLB_2016, hod_EPJC_2015, miao_PLB_2017,schuster_thesis}  as,
\begin{equation}\label{eq_eta_unchrgd}
\eta=\frac{\tau_{gap}}{\tau_{emission}},
\end{equation}
as a figure of merit. $\tau_{gap}$ is the average time interval between the emission of two successive Hawking quanta,
\begin{equation}\label{eq_tgap_unchrgd}
\tau_{gap}=\frac{\omega_{peak}}{P},
\end{equation}
where $P$ is defined in Eq.\eqref{eq_P} and $\omega_{peak} $ is the frequency at which the peak of the power spectrum~(Eq.\eqref{eq_Pl}) occurs, considering complete transmission,  \textit{i.e.,}  the position of the maximum of $\omega^3/\left(e^{\omega/T_{BH}}-1\right)$,
\begin{equation}\label{eq_wpeak_unchrgd}
\omega_{peak}=T_{BH} \left[ 3+ W \left(-3 e^{-3}\right)\right],
\end{equation}
where $W(x)$ is the Lambert $W$-function, defined as \\$W(x) e^{W(x)}=1$. $\tau_{emission}$  is the characteristic time for the emission of individual Hawking quantum and is bounded from below by the localisation time-scale, $\tau_{localisation}$, the characteristic time taken by the emitted wave field  with frequency $\omega_{peak}$ to complete one cycle of oscillation,
\begin{equation}\label{eq_temm_unchrgd}
\tau_{emission} \geq \tau_{localisation}=\frac{2 \pi}{\omega_{peak}}.
\end{equation}

Thus, $\eta\gg 1$ implies that the time gap between the emission of successive Hawking quanta is large compared to the time taken for the emission of individual Hawking quantum, suggesting an extremely sparse Hawking cascade. On the other hand, $\eta\ll 1$ suggests that the Hawking radiation flow is almost continuous. Table \ref{table:1}, \ref{table:2} and \ref{table:3} show the numerical values of  $\eta_{max}=\tau_{gap}/\tau_{localisation}$, $\left(\eta\leq\eta_{max}\right)$ for different values of the scalar and electric charges and the ADM mass.
\begin{table}[!h]
\caption{Numerical values of the dimensionless parameter $\eta_{max}=\tau_{gap}/\tau_{localisation}$ for the $l=1$ mode with $e=0.02$ and $M_{ADM}=1$ for different values of the scalar charge $s$ }
\label{table:1}
\begin{tabular}{cccccc}
\hline\hline  
\rule[-1ex]{0pt}{2.5ex} $s$ & 0.0000 & 0.0002 & 0.0004 & 0.0010 & 0.0050 \\ 
\rule[-1ex]{0pt}{2.5ex} $\eta_{max}$ & 16931.0 & 16927.7 & 16926.1 & 16924.0 & 16921.9 \\
\hline \hline 
\end{tabular} 
\end{table}
\begin{table}[!h]
\caption{Numerical values of the dimensionless parameter $\eta_{max}=\tau_{gap}/\tau_{localisation}$ for the $l=1$ mode with $s=0.001$ and $M_{ADM}=1$ for different values of the electric charge $e$ }
\label{table:2}
\begin{tabular}{cccccc}
\hline \hline 
\rule[-1ex]{0pt}{2.5ex} $e$ & 0.005 & 0.010 & 0.020 & 0.030 & 0.040 \\  
\rule[-1ex]{0pt}{2.5ex} $\eta_{max}$ & 16921.2 & 16921.4 & 16924.0 & 16931.6 & 16945.3 \\ 
\hline \hline 
\end{tabular} 
\end{table}
\begin{table}[!h]
\caption{Numerical values of the dimensionless parameter $\eta_{max}=\tau_{gap}/\tau_{localisation}$ for the $l=1$ mode with $e=0.02$ and $s=0.001$ for different values of $M_{ADM}$ }
\label{table:3}
\begin{tabular}{cccccc}
	\hline \hline 
	\rule[-1ex]{0pt}{2.5ex} $M_{ADM}$ & 1.00 & 0.80 & 0.50 & 0.20 & 0.15 \\  
	\rule[-1ex]{0pt}{2.5ex} $\eta_{max}$ & 16924 & 16925.6 & 16932.4 & 16991.5 & 17046.6 \\ 
	\hline \hline 
\end{tabular} 
\end{table}
The high values of the dimensionless parameter suggest that the Hawking cascade of massless uncharged scalar quanta from the scalar-hairy RN black hole is extremely sparse. Table~\ref{table:1} and Table~\ref{table:2} show that with the increase of scalar charge $s$, the sparsity decreases as opposed to the enhancement of the sparsity of the Hawking cascade with the electric charge $e$. Table~\ref{table:3} shows that for fixed $e$ and $s$, the sparsity of the Hawking radiation cascade increases with the decrease in the ADM mass of the black hole. Though these variations are steady and monotonic, the variations are rather small.


\section{Summary and Discussion }\label{sec:summary}

Despite the rather modest appearance of the scalar hair in the scalar-hairy-Reissner-Nordstr\"{o}m black hole (see Eq.\eqref{eq_metric}) recent studies \cite{astorino_PRD_2013, chowdhury_EPJC_2018, chowdhury_EPJC_2019} have shown that the presence of this additional scalar field greatly modifies the known physics of the standard Reissner-Nordstr\"{o}m black hole.  In the present study, we analyzed the greybody factor and  also estimated the dimensionless parameter, $\eta=\tau_{gap}/\tau_{emission}$, to study the sparsity of Hawking radiation from the scalar-hairy-Reissner-Nordstr\"{o}m black hole. We considered the emission of massless uncharged scalar quanta for this purpose.

We observed that the black hole scalar and the electric charges oppositely affect the greybody factor.  Increasing the scalar charge increases the greybody factor, whereas, increasing the electric charge has the effect of lowering the greybody factor.  For a given electric charge, the total Hawking radiation power emitted in each mode decreases with the scalar charge. The peak of the power spectrum also diminishes and shifts towards lower frequencies. This in-turn reduces the sparsity of Hawking radiation flow with the scalar charge. However,  increasing the electric charge increases the power of Hawking radiation. The peak value of the emitted power spectrum also increases and shifts towards higher frequencies and the Hawking flux becomes even more sparse.  
 
As the black hole continues to Hawking radiate, its ADM mass decreases. The lowering of the ADM mass, on one hand, raises the black hole temperature $T_{BH}$ (see Eq.\eqref{eq_TBH}) and enhances the Hawking emission power, while on the other hand, it reduces the grey body factor and  increases the sparsity of the Hawking radiation cascade. However, for the emission of uncharged particles, the ADM mass of the black hole cannot reduce to zero since it is bounded by the mass of the extremal black hole, $M_{ADM}\geq \frac{e^2}{\sqrt{e^2+s}}$. 
 
 It is also interesting to note that if quantum gravity effects result in the formation of a black hole remnant at the end stages of Hawking radiation\cite{chen_PR_2015, ong_JHEP_2018, ong_PLB_2018, gohar_PRD_2018, gohar_IJMPD_2018}, then, Eq.\eqref{eq_MADM} suggests that for non-zero scalar charge, the remnant has to be electrically charged. A detailed study of such a remnant is outside the scope of this article and will be considered in a separate work.

Both the greybody factor and the sparsity are considered to be important characteristics of Hawking radiation, which should help understanding the mechanism of the quantum processes involved. As the modes of scalar fields are instrumental in the formulation of Hawking radiation, these investigations are likely to be extremely relevant.

\bibliographystyle{elsarticle-num}
\bibliography{ref3}
\end{document}